\documentclass[10pt,twocolumn]{article}
\usepackage[margin=1in]{geometry}
\usepackage{comment}
\usepackage{ifthen}
\usepackage{graphicx}
\PassOptionsToPackage{hyphens}{url}\usepackage{hyperref}
\usepackage{times}
\usepackage{amsmath}
\usepackage{amssymb}
\usepackage[bottom]{footmisc}

\title{Intrusion Detection and Ubiquitous Host to Host Encryption}
\author{Aaron Gibson and Hamilton Scott Clouse \\
aaron@the-gibsons.org, hsclouse@ieee.org}
\date{December 4, 2011}
\begin{document}
\maketitle

\begin{abstract}
\textit{
Growing concern for individual privacy, driven by an increased public awareness of the degree to which many of our electronic activities are tracked by interested third parties (e.g. Google knows what I am thinking before I finish entering my search query), is driving the development anonymizing technologies (e.g. Tor). The coming mass migration to IPv6 as the primary transport of Internet traffic promises to make one such technology, end-to-end host based encryption, more readily available to the average user. In a world where end-to-end encryption is ubiquitous, what can replace the existing models for network intrusion detection? How can network administrators and operators, responsible for securing networks against hostile activity, protect a network they cannot see? In an encrypted world, signature based event detection is unlikely to prove useful. In order to secure a network in such an environment, without trampling the privacy afforded to users by end-to-end encryption, our threat detection model needs to evolve from signature based detection to a heuristic model that flags deviations from normal network-wide behavior for further investigation. In this paper we present such a heuristic model and test its effectiveness for detecting intrusions in an entirely encrypted network environment. Our results demonstrate the network intrusion detection system's ability to monitor a network carrying only host-to-host encrypted traffic. This work indicates that a broad perspective change is required. Network security models need to evolve from endeavoring to define attack signatures to describing what the network looks like under normal conditions and searching for deviations from the norm.
}
\end{abstract}

\section{Introduction}
In computer networks, specifically Internet Protocol (IP) networks, hosts or devices are identified by unique identifiers. These labels, as of the writing of this paper, are predominantly of a form dictated by the IPv4 standard. In this standard a network connected device's is described by a sequence of four numbers mapping its connectivity to the network. For instance, as in sub-netting practices, the last number of the identification sequence indicates the device as a member of a small section of the whole IP network (a subnet). The whole sequence is referred to as an IP address and the individual numbers are ``octets," so called since they can take on values in the range $[0,2^8]$. Since this addressing scheme only affords 4.3 billion unique identifiers, a single IP network would be limited to that many devices.

Appropriately so, the most well-known IP network is the Internet. Since its inception it has grown exponentially into the ubiquitous connectivity medium that it is today. Furthermore, the worldwide explosion of the popularity of wired, wireless and mobile connected devices has exhausted the 4.3 billion unique identifiers. Historically, solutions to the limitations of addressing Internet connected devices have been implemented with success, e.g. network address translation (NAT) and variable-length subnet masks, as referenced previously. Such practices have directed the evolution of common network topologies towards those as illustrated in Figure \ref{fig:Figure 1}. Confidential communication between devices on separate local area networks (LAN's) over the Internet, has thus been mostly confined to IP tunneling and virtual private network (VPN) configurations.

The transition to the more expansive addressing scheme IPv6 generates almost 100 orders of magnitude more available unique identifiers, thus negating the need for schemes such as NAT to separate potentially identical addresses. If the current Internet were fully realized converted to the IPv6 standard, every connected device would have its own unique public address. In this setting direct confidential communication between devices is viable, through host-to-host encryption, since no two hosts will have the same address. While this is a boon to confidentiality goals, it is a detriment to current network security measures: \cite{AlSadeh2011},\cite{4781968},\cite{Zagar2007425}, and \cite{2461970320070402}.

Intrusion detection systems (IDS), and particularly network IDS (NIDS) rely on information extracted from packet traffic on the LAN to recognize breaches of security policies and issue alerts to administrators or network management software. Current NIDS utilize much data available in the packet, including the protocols and ports employed during a conversation, to compute statistics or generate patterns for analysis: \cite{Marchette2001},\cite{Abad2004},\cite{Locatelli2004}, and \cite{Moya2008}. Implementation of host-to-host encryption would prohibit a NIDS access to such information, limiting it to only the information in the packet headers, e.g. source and destination addresses, payload length, and time to live. Traffic analysis on this limited data is difficult and research in this area is ongoing: \cite{Pilgermann_anonymizingdata},\cite{springerlink:10.1007/978-3-642-15512-3_35},\cite{5635013}, and \cite{Jarvinen2011}. Table \ref{tab:results} serves to illustrate the typical intrusion detection rates for current NIDS.

\begin{table*}[]
\centering
\caption{Probability of detection $>40\%$ for NIDS in various OS/attack pairs.}
\label{tab:results}
\setlength\tabcolsep{2pt}
\begin{tabular}{l|l|l|l|l|l}
\multicolumn{1}{r|}{\textbf{Pr\{detect\}}} & DoS & Probe & R2L & U2R & Data \\ \hline
Solaris & \begin{tabular}[t]{lr}Expert-1 & 63\%\\ Expert-2 & 53\%\end{tabular} & \begin{tabular}[t]{lr}Expert-2 & 60\%\\ Expert-3 & 50\%\end{tabular} & \begin{tabular}[t]{lr}Expert-1 & 50\%\\ Forensics & 50\%\end{tabular} & \begin{tabular}[t]{lr}Expert-1 & 100\%\\ Expert-2 & 100\%\\ Anomaly & 100\%\\ Forensics & 73\%\end{tabular} & \begin{tabular}[t]{lr}Expert-2 & 100\%\\ Forensics & 83\%\end{tabular} \\ \hline
NT & \begin{tabular}[t]{lr}Expert-1 & 69\%\\ Expert-2 & 69\%\end{tabular} & \begin{tabular}[t]{lr}Expert-1 & 80\%\\ Expert-2 & 60\%\end{tabular} &  &  &  \\ \hline
SunOS & \begin{tabular}[t]{lr}Dmine & 88\%\\ Expert-1 & 63\%\\ Expert-2 & 50\%\end{tabular} & \begin{tabular}[t]{lr}Pclassify & 60\%\end{tabular} & \begin{tabular}[t]{ll}Expert-2 & 67\%\end{tabular} &  &  \\ \hline
Linux & \begin{tabular}[t]{lr}Dmine & 74\%\\ Expert-1 & 84\%\\ Expert-2 & 68\%\end{tabular} & \begin{tabular}[t]{lr}Dmine & 50\%\\ Expert-3 & 60\%\end{tabular} & \begin{tabular}[t]{lr}Expert-1 & 64\%\\ Expert-2 & 44\%\end{tabular} &  &  \\ \hline
All &  & \begin{tabular}[t]{lr}Expert-1 & 46\%\end{tabular} &  &  & 
\end{tabular}
\end{table*}


As can be seen, few of the techniques perform with high detection percentages and none perform well for all OS/attack vector pairings. These techniques are largely separable into two classes: expert/heuristic systems and learning/data driven systems. The two approaches, while independently feasible for some applications, are not broadly applicable to the IDS task. In this effort we examine these two approaches via two pairings: expert inspired statistical analysis and data driven heuristic definitions. The paper is organized thus: Section \ref{sec:relWork} describes past and current efforts on this issue, Section \ref{sec:Needle} expands on the problem and our efforts are discussed in Section \ref{sec:Work}.

\section{Related Work}\label{sec:relWork}
\subsection{Network Monitoring and Network Data Confidentiality}
\begin{figure}[ht!]
\center{ \includegraphics[width=80mm]{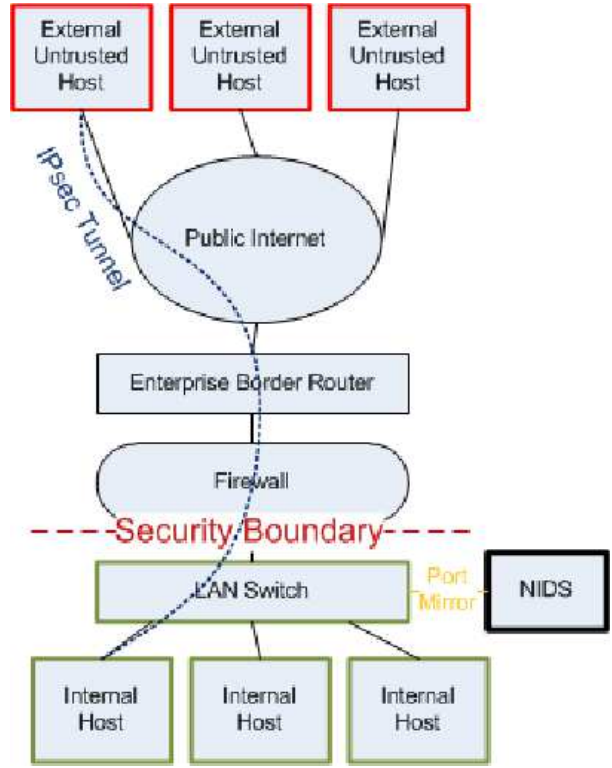}}
\caption{\label{fig:Figure 1}Common Enterprise Network Topology}
\end{figure}

The fundamental challenge of network intrusion detection in the presence of ubiquitous encryption is the inherent tension between competing security concerns: on the one hand, enterprise network users maintain a legitimate desire to protect the confidentiality of their communications over the network while on the other hand, network administrators are tasked with the responsibility of ensuring that internal network and computing resources are not compromised by the intrusion of unauthorized activity. Figure  \ref{fig:Figure 1} depicts the challenge posed to an enterprise network administrator in the presence of encrypted user traffic through the network.  The NIDS in the figure will not have access to information in the encrypted payload of the packets transported via the IPsec tunnel.  In recent years the desire to protect both the confidentiality of data in transit and corporate resources from intrusion has intensified. Applications and protocols offering transparent encryption of user traffic have become increasingly available.  Because RFC4301 defines the implementation of IPsec as mandatory for all IPv6 nodes, the migration to IPv6 as the primary network transport protocol promises to make transparent encryption an even more ubiquitous reality. At the same time, occurrences of malicious cyber activity and hostile intrusions into private networks and computing resources are on the rise.  

\subsection{Prior Approaches to the Problem}
A number of different approaches have been proposed in an attempt to resolve this tension.  Schaffrath \cite{Schaffrath2008}, and Studer et al.\cite{Studer_tuningintrusion}, both attempt to address the problem by separating the encryption of the IPSec header and payload, each with a distinct key, and augmenting the Network Intrusion Detection system (NIDS) with Host-based Intrusion detection systems (HIDS). While the ``Two-Key IPSec" approach does maintain the confidentiality of user data it also requires the modification of the IPSec protocol which would represent a substantial overhaul of widely deployed software, much of which is built-in at the operating system level, and fails to address how preventing the user from encrypting traffic using traditional (single key) IPSec would be enforced.

Goh et al., in a series of papers exploring the problem of intrusion detection in encrypted networks (\cite{Goh2010},\cite{GohEtal2010},\cite{GohEtal2010a} and \cite{Goh2009}) develop a framework that introduces a secret sharing scheme together with a number of randomized network proxies to ensure that both the intended receiver and the NIDS obtain a decryptable copy of the user traffic. However, this solution requires a significant increase in the amount of traffic on the local area network (LAN), at least doubling the traffic volume. The solution also requires modification of every host on the network.

Irwin \cite{Irwin05} proposes a central vault server for the storage of cipher keys for all encrypted services provided by the monitored network. In his design the IDS is able to retrieve the keys necessary to decrypt traffic for analysis from the central vault server. It would seem that this creates a significant vulnerability to the confidentiality of all the services provided on the network. One successful intrusion to the central vault server could render all encryption on the network useless.

\subsection{Communities of Interest and Flow Analysis}
In this paper we investigate whether enough information is contained within the encrypted data flows for a NIDS to detect malicious behavior without decrypting the network traffic.  Past research on detecting communities of interest and normal traffic patterns that do not rely on knowledge of the content of the traffic can be leveraged to resolve the problem of intrusion detection in encrypted networks.

Aiello et al., provide a particularly relevant discussion\cite{Aiello2005}. This paper explores the ability of community of interest (COI) analysis to characterize and predict the behavior of hosts within a data network. If normal behavior can be characterized, so too can anomalies. The methodology of evaluating various aspects of COIs of hosts within an IP network, provided in this paper, offers a promising avenue to approach the problem of intrusion detection on an encrypted network.

Much additional research has been done on the extent to which analysis of NetFlow records can provide assistance in both defining the normal state of a network and in detecting abnormalities.  Bin et al.\cite{Bin20081074},  propose ``a real-time anomalous traffic monitoring module of a NetFlow monitoring system with a stable matching pattern algorithm and two traffic statistic based intrusion detection algorithms - one algorithm is based on variance similarity and the other based on Euclidean distance to detect worms and other malicious attacks." Their system demonstrates the usefulness of flow record analysis in intrusion detection. While it is not written for traffic flows in an encrypted environment, the work that they have done does provide a basis from which to begin the flow analysis of encrypted traffic.  

\section{Finding the Needle in a Needle Stack}\label{sec:Needle}
The challenge of generalizing flow analysis to be relevant to the more limited data set available in the presence of encrypted flows is not trivial.  Even the definition of a flow must be revisited in the presence of IPsec.  A network flow has been traditionally defined by at least a 4-tuple compromised of the source IP address, destination IP address, source port and destination port.  In the presence of IPsec the source and destination ports are not available for analysis.  Even the transport protocol is hidden by the encryption, as the protocol in the IP header of an IPsec flow is 50 (the protocol number for IPsec itself).

Figure 2 depicts the information that is contained in IPv4 and IPv6 headers.  This is the only information that is available to a NIDS for analysis in the presence of IPsec.  And while the IP headers are available for analysis, some of their fields are trivially uninteresting.  As previously noted, the IPv4 Protocol or IPv6 Next Header field will always be equal to the IPsec protocol number.  The Version field provides no additional information for analysis.  Neither the IPv4 Type of Service nor the IPv6 Traffic Class fields have been used widely enough to provide data of interest.  As such, in the presence of IPsec the NIDS is limited to analysis of the source and destination addresses, packet size and Time To Live in IPv4 or Hop Limit in IPv6 for its characterization of traffic. 

Our goal was to  use the information that is available for the analysis of traffic patterns to create a network intrusion system that could be trained to detect anomalous malicious behavior without the need to decrypt network traffic. While we were unable to create such a system we were able to identify some trends that point in the direction of a solution.  In this paper we describe that data set used as the basis for our analysis, the techniques used to analyze the data and the results of that analysis.  We took a parallel approach to the analysis of the data, separately applying machine learning techniques to the data set as an abstract and statistical analysis of the data based on characteristics likely to differentiate normal and attack traffic.

\begin{figure}[ht!]
\center{ \includegraphics[width=80mm]{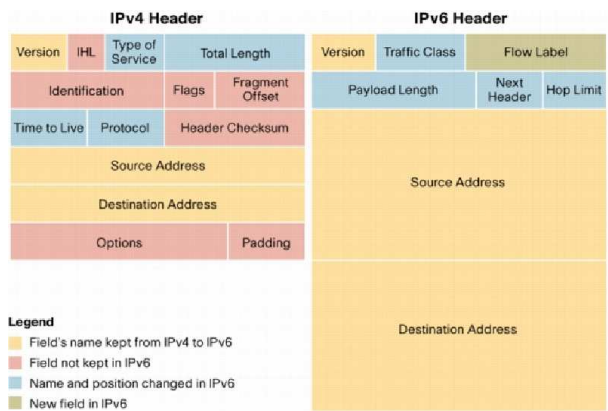}}
\caption{\label{fig:Figure 2}IPv4 and IPv6 Header Fields}
\end{figure}

\section{Data Analysis and Characterization of Network Traffic}\label{sec:Work}

\subsection{The 1998 DARPA IDS Dataset}
While the consensus of the network security community is that the DARPA IDS data sets and the derived 1999 KDD Cup data are fundamentally broken, even the harshest critics generally acknowledge that there is no superior publicly available data set relevant to the evaluation of IDS.  With that in mind, we acknowledge the shortcomings of the source data used for our investigation and concede that any definitive conclusion would require testing on additional data sets.  However, as the nature of our work is only to point towards avenues that appear to be fruitful in analyzing encrypted traffic to identify intrusions, we do not believe that the flaws in the source data detract from the claims that we make in this paper.  As a matter of fact, the flaws in the source data rather make us more hopeful that network intrusion detection is possible in the presence of ubiquitous encryption.  Because one of the published flaws with the DARPA data set was the irregularity in the Time To Live that allowed trivial differentiation of attack traffic, we limited our analysis to packet length and IP addresses.  Removing this variable from our analysis only made the differentiation of attack traffic more difficult.

\subsection{Analysis of Characteristics Likely to Differentiate Attacks}
\subsubsection{Data Analysis Methodology}
The analysis of the data set based on characteristics likely to differentiate attack traffic was performed on an Ubuntu Linux server running MySQL 5.1.58.  The Linux wget utility was used to download the daily archives from the MIT Lincoln Laboratories website\footnote{http://www.ll.mit.edu/mission/communications/ist/corpora/ideval/data/1998data.html}.  Extraction of the archive files was scripted using Perl.  Upon extraction the only file used for import into the MySQL database was the tcpdump file.  The Linux utility tshark was used to extract the relevant fields from the packet header using the following command:
\begin{verbatim}
tshark -n -T fields \
-e ip.src \
-e ip.dst \
-e ip.len \
-e ip.ttl \
-e frame.time_epoch \
-r tcpdump > headers
\end{verbatim}
\vspace{5 mm}

Database tables in the following format were created to store the packet header data to be analyzed:
\scriptsize
\begin{verbatim}
+--------+---------------------+------+-----+---------+
| Field  | Type                | Null | Key | Default |
+--------+---------------------+------+-----+---------+
| src_ip | bigint(20) unsigned | YES  | MUL | NULL    |
| dst_ip | bigint(20) unsigned | YES  | MUL | NULL    |
| length | mediumint(9)        | YES  |     | NULL    |
| ttl    | smallint(6)         | YES  |     | NULL    |
| ts     | decimal(15,6)       | YES  | MUL | NULL    |
| attack | tinyint(3) unsigned | NO   | MUL | 0       |
+--------+---------------------+------+-----+---------+
\end{verbatim}
\normalsize
The labeling information was also obtained from the Lincoln Laboratories website\footnote{http://www.ll.mit.edu/mission/communications/ist/corpora/ideval/docs/attacks.html}.  Because the attack instances were not labeled on a per packet basis, a Perl script was used to read the attack schedule file and populate the \verb+attack+ field in the database tables.  The logic for labeling the packets in the database was to assign a value of 1 to source and destination IP pairs matching those in the attack list near the time of the attack in the schedule.  A value of 2 was assigned to database entries where the source and destination IP addresses were reversed.  This resulted in three categories of labels.  Normal traffic was labeled 0.  Attack traffic was labeled 1. Replies to attack traffic were labeled 2.  Importing and labeling the data in this manner allowed a number of analyses to be run comparing the statistical characteristics of the available parameters.

As was discussed earlier, because source and destination port are not available for analysis, an alternate method of distinguishing a flow was required.  For the purposes of this analysis a flow was defined as a grouping of packets sent from A to B within a specified time slice.

\subsubsection{Analysis of Packet Size Variation} 
\label{sec:Analysis of Packet Size Variation}
Given the limited set of variables available to differentiate attack traffic in the presence of IPsec encrypted traffic, a careful
reasoning concerning likely differences is required.  One generalization that can be made regarding attack traffic is that it is much more likely to be automated than normal traffic. This recognition led to the inference that it may well be the case that attack traffic is more uniformly sized within a given flow than normally occurring network traffic.

The statistical examination of the data confirms this inference.   While there did not appear to be any discernible difference in average size of packets in an attack than in a normal traffic flow, the standard deviation of the size of packets in an attack tended to be be much smaller than that seen for normal traffic.  A frequency distribution of the standard deviations of packets size in a given flow for both normal and attack traffic follows:
\begin{figure}[ht!]
\center{ \includegraphics[width=80mm]{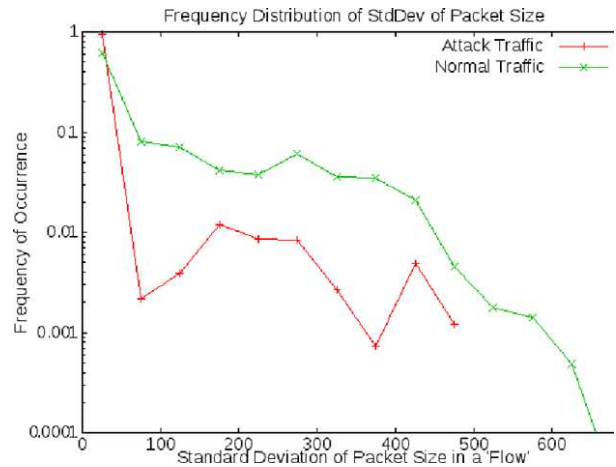}}
\caption{\label{fig:Figure 3}Degree of Packet Size Variation within a Flow}
\end{figure}
As can be seen in the preceding frequency distribution, standard deviations of 50 bytes or smaller occru in only 60\% of the normal traffic flows and standard deviations in packet size as large as 400 bytes occur in nearly 5\% of normal flows.  Conversely the stand deviation in packet size for attack flows is 50 bytes or less for more than 95\% of all flows.

\subsubsection{Community of Interest Based Analysis}
One analysis technique that seemed likely to be useful in identifying attack instances was the identification of communities of interest derived from typical host communication patterns.  Aiello et al., provide an analysis that suggests "some stability in the COI for the community as a whole" \cite{Aiello2005}.  In our analysis we employed a variant of of what Aiello et al., called a frequency based community of interest.  Based on the definition offered by Aiello et al. of a frequency based community of interest:\cite{Aiello2005}
\begin{quote}
A host is considered to be part of the COI of a target-host, if the targethost interacts with it at least once every small time-period Z (the bin-size) within some larger time period of interest Y.
\end{quote}
In attempting to apply this definition to the DARPA data set, it was determined that one could more effectively capture membership in a legitimate community of interest by both expanding the bin-size and looking at the percentage of bins in which the interaction occurred rather than requiring interaction occur in \textit{every} bin.  This analysis is based on the likelihood that communications between a internal host and an attack source would be relatively infrequent given the infrequency of  occurrence of attack traffic.  However, the following graph showing the frequency distribution of ‘flows’ from A to B, does not appear show any appreciable difference in distribution of attack pairs and normal pairs.  For this analysis “Attack Traffic” was defined as any ‘flow’ A to B such that at some point in the data set an attack occurred form A to B.  
\begin{figure}[ht!]
\center{ \includegraphics[width=80mm]{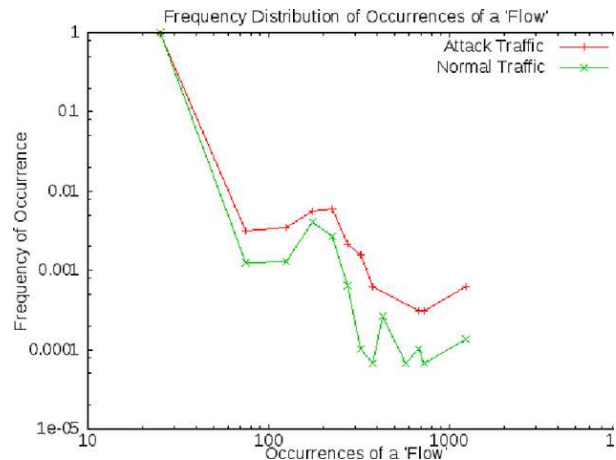}}
\caption{\label{fig:Figure 4}Frequency of Occurrence of a Flow}
\end{figure}

This result was unexpected and further analysis of the dataset showed that a large number of attacks were sourced from IP addresses that were also used for a relatively high frequency of normal traffic flows.  It would be worth investigating whether this is common in live network traffic or is unique to this dataset.

In addition to analyzing the regularity with which a flow from A to B is repeated over a longer time period, we attempted to find ways to characterize the flows in ways that would allow us to differentiate between normal and attack traffic.  One such attempt, the characterization of packet size variation, was discussed in section~\ref{sec:Analysis of Packet Size Variation}.  Our analysis of the total number of packets transmitted in a flow from A to B showed some distinguishing characteristics of attack traffic.  Unfortunately, the distinguishing characteristics seem only to apply to trivial cases.  You hardly need a sophisticated intrusion detection engine to notify that you are currently in the midst of a Denial of Service attack.
\begin{figure}[ht!]
\center{ \includegraphics[width=80mm]{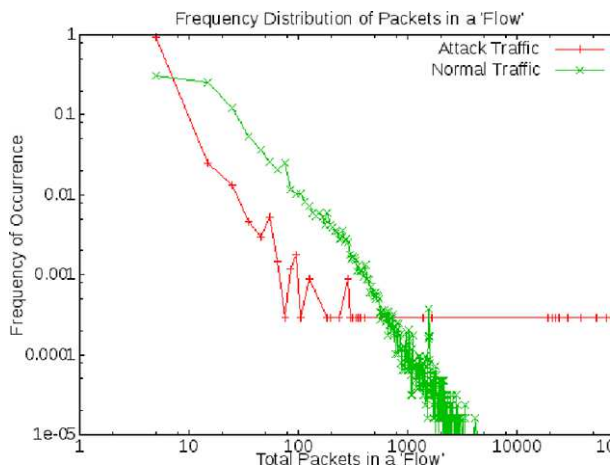}}
\caption{\label{fig:Figure 5}Number of Packets within a Flow}
\end{figure}

As depicted in Figure~\ref{fig:Figure 5} instances of receipt of hundreds of thousands of packets in a single flow seem to be quite unique to attack traffic.  However, aside the obvious Denial of Service attacks, the frequency distribution of packets sent in a flow are rather similar for both attack and normal traffic.  The curve for the attack traffic is decidedly less smooth, but this likely a function of the smaller sample size more than an underlying difference in actual traffic pattern.
\subsection{Inferring Intuition from Data Analysis}
The other point of view from which the problem was approached was that of data driven learning. By studying the intrinsic nature of the data, it might be possible to glean insight about natural separations in the data, i.e. discerning between ``normal" and ``malicious" traffic. Analysis began by extracting the four-dimensional feature vector $feat_i$ for each packet $i$, as given in equation \ref{eqn:featureVec}. 
\begin{equation}\label{eqn:featureVec}
feat_i=\begin{bmatrix}
time_i \\
src_i \\
dest_i \\
len_i \end{bmatrix}
\end{equation}
\subsection{Initial Exploration}
\begin{figure*}[h]
\center{ \includegraphics[width=0.8\textwidth]{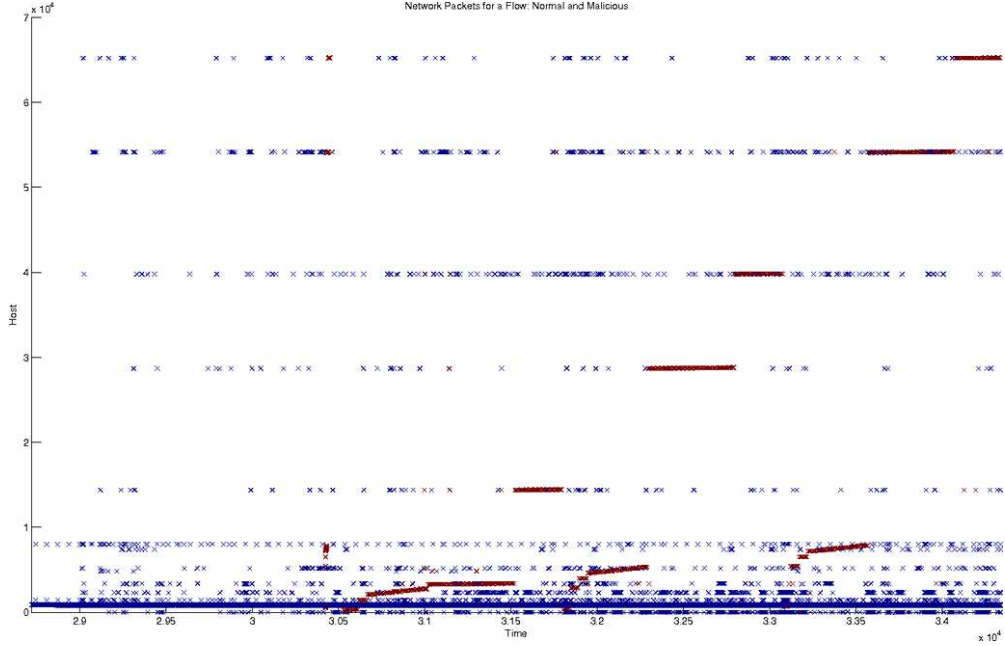}}
\caption{\label{fig:hostsTime}Network packets (normal and malicious) as a function of time for each host.}
\end{figure*}

An initial exploratory effort led to the production of the graph in Figure \ref{fig:hostsTime}, a plot of a series (or flow) of packets for every host on the network. Each point represents a packet and the color represents the nature of that packet (blue=``normal", red=``malicious"). It is evident that patterns in the attack packets do exist, but it also evident that nothing about a single packet would be useful in recognizing an attack vector. So, more information must be assumed, e.g. windowing in time or structure of the data space. Since the statistical/parallel effort focused on the time windowing, here the possibility of structure within the data was considered.

By the assumption that there is underlying structure/patterns in the data, the problem can be formulated as a manifold learning task. Many algorithms are available to discover the salient topology of such an unknown manifold. Of particular interest is a newer algorithm: Diffusion Maps. This approach utilizes a random walk on a graph constructed from the feature vectors. In this way, the true structure of the data precipitates the mapping. The resultant mapping then represents the data in Euclidean space isomorphically, i.e. maintaining the relative structure between data points. In this form, many common geometric and statistical tools become viable that would not have been if the original non-flat feature space were exploited.

\subsubsection{Diffusion Maps}
The Diffusion Map technique is a non-linear dimension reduction technique introduced in \cite{DiffMaps,DataFuse} and is referred to in this paper as the Diffusion Map. The Diffusion Map first embeds raw data into a spectral graph framework. The method for defining the 'nodes' and 'edges' that comprise the graph is application specific. Edges provide a measure of 'similarity' between nodes. This measure should have these two properties: symmetry and non-negativity. That is, the similarity between two nodes should be the same regardless of which node is used as a reference. The non-negativity is straightforward. The name "Diffusion" is a reference to the process of heat diffusing through a medium\cite{website:NonLinDimRed}. Similar to the model of heat diffusion, in this algorithm weights are assigned to the edges of the graph, as related to a reference node, by a random walk on the graph with a distribution that diminishes the farther away from the reference node the walk progresses.

\begin{figure}
\centering
\includegraphics[width=0.2\textwidth]{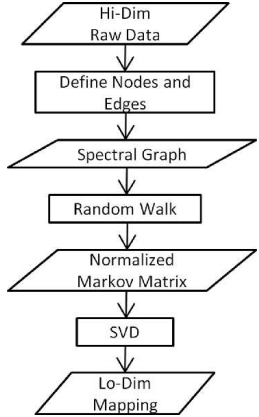}
\caption{Flow chart representing the Diffusion Map algorithm.}
\label{fig:DiffMapFlow}
\end{figure}

With these weights considered as probabilities, one can form a transition probability matrix, a Markov Matrix, representing the data as it lies on the manifold. When the assumed existence of the manifold holds, the eigenvectors of the normalized Markov matrices embed the graph into a Euclidean space. The algorithm for the Diffusion Map used in this paper is succinctly outlined in \cite{DiffMapRadar}. The overall notion of the technique is depicted in Figure \ref{fig:DiffMapFlow}. We refer the reader to \cite{DiffMaps} and \cite{DataFuse} for the complete details of the Diffusion Map method.

We implemented the Diffusion Map technique using the algorithm detailed in \cite{DiffMapRadar}, choosing the weights w(i,j,), using a Gaussian distribution, described in Equation \ref{eqn:Gauss} below, where the indicated norm is the Euclidean, $L^2$, distance as also chosen in \cite{DataFuse} and \cite{DiffMapRadar}. 
\begin{equation}
\label{eqn:Gauss}
e^{\frac{-\|x_i-x_j\|^2}{\epsilon}}
\end{equation}
By appropriately normalizing these edge weights, the transition probabilities for a random walk on the graph were determined. These transition probabilities were then formulated as a Markov matrix. Thus, by determining the eigenvalues and eigenvectors of this matrix one can embed the graph into Euclidean space using the Diffusion Map given by Equation \ref{eqn:DiffMap} to achieve dimension reduction to a space intrinsic to the data.
\begin{samepage}
\begin{equation}
\label{eqn:DiffMap}
\Psi_t:x\mapsto(\lambda_2^t\psi_2(x),\lambda_3^t\psi_3(x),\ldots,\lambda_m^t\psi_m(x))^T
\end{equation}
Here $m(t)$ is the number of terms retained to define the diffusion map and embed the data into the Euclidean space $R^{m(t)}$, $\lambda_i$ are the eigenvalues, $\psi_i$ are the eigenvectors, and $t$ is the exponential of the resulting eigenvalues. The difficult part of using the Diffusion Map technique is finding the appropriate value for $\epsilon$, the diffusion-distance tuning parameter and $t$, the exponent of the eigenvalues. As $\epsilon$ increases, the edge weight increases, and as $t$ increases, the spectrum decays at a greater rate\cite{DiffMapRadar}.
\end{samepage}

\begin{figure*}
\centering
\includegraphics[width=160mm]{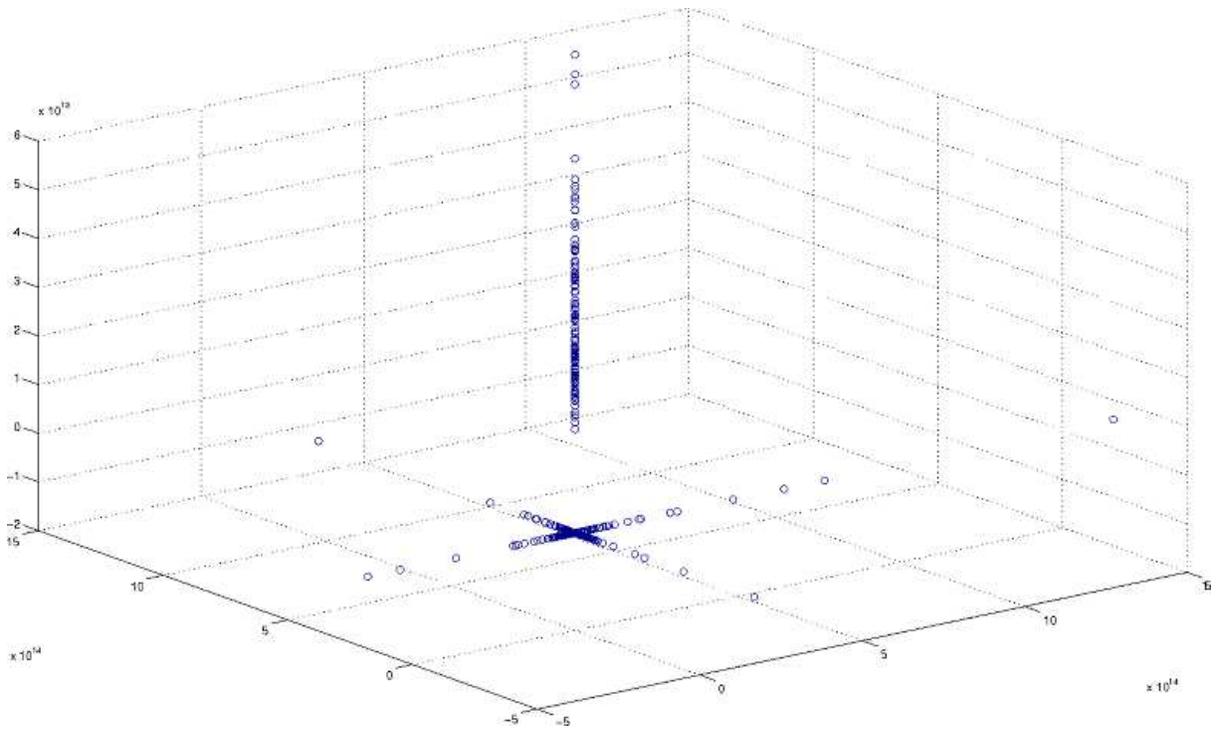}
\caption{Three-dimensional mapping resulting from the Diffusion Map algorithm. Note the clear orthogonality in the data.}
\label{fig:DiffMapping}
\end{figure*}

The resulting three-dimensional mapping of the data (Figure \ref{fig:DiffMapping} did indicated that a very clear structure exists. However, the same coloring scheme utilized in Figure \ref{fig:hostsTime} was employed and the ``normal" packets completely occlude the ``malicious" packets. So, while the data has an intrinsic structure that is not captured directly by the feature vector, it is not obvious that the structure is related to the intent behind the packets.
\subsection{Unsupervised Clustering Classification: K-Means}
The last method implemented in studying the intrinsic nature of the data was a method popular in data-mining problems: clustering. These techniques force a separation onto the data set of a give order. The K-Means algorithm is simple: initialize with K random points in the feature space then 1)assign each observation to the cluster with the nearest mean and 2)calculate the new clusters' means. This process is iterated until cluster assignments cease to change. Since the desired distinction was ``normal" vs. ``malicious" a K-value of two was set; this ensured that the data was forced to separate into two clusters. This method does not lend itself to visualization, but rather classification. The labels assigned to the packets via the clustering technique were compared to the truth labels provided with the data sets with the following results (Table \ref{table:KMeans}).
\begin{table}
\centering
\caption{K-Means Clustering Accuracy}
\label{table:KMeans}
\begin{tabular}{|l|c|}
\hline
True Positive & $83.30\%$\\
\hline
False Positive & $3.12\%$\\
\hline
True Negative & $0.46\%$\\
\hline
False Negative & $13.30\%$\\
\hline
\end{tabular}
\end{table}
This approach, while showing a high accuracy, also exhibited a higher false negative rate. So, clustering, as with Diffusion maps, seem to arrive at the intrinsic structure of the data, but more information is necessary to ascertain the nature of that structure.

\section{Conclusion and Future Work}

Both of our parallel efforts to analyze the structure and distinguishing characteristics of the portion of the DARPA 1998 IDS dataset that would be available in the presnece of encrypted flows lead us to conclude that Intrusion Detection Analysis limited to the IP header can be a significant contributing factor in detecting attack traffic.  The application of the manifold learning technique (an unsupervised/unlabeled approach), Diffusion Maps, was fruitful in that some intrinsic structure of the data was indicated given only the four variables: time, src, dest, and length. However, due to the non-linear nature of the algorithm, an inverse mapping to determine the crux of that structure is not possible. To ascertain the structure, an unsupervised clustering scheme was employed. This approach, while it illustrated a separation in the data (possibly artificially), it was not the separation between "normal" and "malicious" packets that was desired. Further investigation of this relationship is warranted; specifically with a manifold learning algorithm that exhibits a more tractable inverse.  The statistical analysis of normal and attack traffic characteristics likely to differentiate the traffic types did provide evidence that attack flows are more uniform in packet size.  While the community of interest analysis based on the previous work of Aiello et al., \cite{Aiello2005} did not result in the anticipated level of success, additional work to determine if this shortcoming was related to peculiarities in the DARPA dataset is warranted.

Regardless, the results of our statistical analysis and the application of the manifold learning technique lead us to conclude that it is not possible to differentiate between attack traffic and normal traffic on a per packet basis without incorporating additional information.  However, it does appear likely that in many cases relevant information could be fed into the system.  In most enterprise networks significant knowledge about both the logical topology and functional requirements for hosts on the network are readily available.  Given the availability of this information, less naive metrics could be utilized to account for/include knowledge concerning the topology of the network to directly influence the structure of the learned manifold.  Expected patterns of traffic could be more readily identified based on a hosts known functional role within the organization.  One might reasonably expect that what constitutes a normal flow for a corporate mail server would be dramatically different from a normal flow for a user workstation.  With the ability to profile normal flows customized to specific IP addresses or IP ranges the intrusion detection system would have significantly greater information form which to begin the analysis of the limited set of available variables. Additional investigation of how best to group packets sequences into a flow may also lead to improved performance in the ability of the community of interest analysis in distinguishing between normal and attack traffic.  

While we were not able to develop a system that could learn to distinguish attack traffic from normal traffic relying on analysis of only the IP addresses, packet sizes and arrival times we did obtain results that give some hints to where future work in the area is likely to be fruitful.
\bibliographystyle{abbrv}
\bibliography{Clouse_Gibson}

\end{document}